\documentclass[sigconf,natbib=false,balance=false]{acmart}
\usepackage{soul}
\usepackage{url}
\usepackage{amsthm}
\usepackage[ruled, lined, linesnumbered, commentsnumbered, longend]{algorithm2e}

\usepackage{lipsum} 
\usepackage{caption}
\usepackage{subcaption}

\usepackage{multirow}

\usepackage{xspace}
\newcommand{\proj}{FedKEM\xspace} 

\RequirePackage[
  datamodel=acmdatamodel,
  style=acmnumeric,
  ]{biblatex}

\copyrightyear{2023}
\acmYear{2023}
\setcopyright{rightsretained}
\acmConference[SC-W 2023]{Workshops of The International Conference on
High Performance Computing, Network, Storage, and Analysis}{November
12--17, 2023}{Denver, CO, USA}
\acmBooktitle{Workshops of The International Conference on High Performance
Computing, Network, Storage, and Analysis (SC-W 2023), November 12--17,
2023, Denver, CO, USA}
\acmDOI{10.1145/3624062.3626325}
\acmISBN{979-8-4007-0785-8/23/11}

\begin{document}
\pagestyle{plain}

\title{Enhancing Heterogeneous Federated Learning with Knowledge Extraction and Multi-Model Fusion}


\author{Duy Phuong Nguyen}
\authornote{Both authors contributed equally to this research.}
\affiliation{%
  \institution{Iowa State University}
  \city{Ames, IA}
  \country{USA}}
\email{dphuong@iastate.edu}

\author{Sixing Yu}
\authornotemark[1]
\affiliation{%
  \institution{Iowa State University}
  \city{Ames, IA}
  \country{USA}}
\email{yusx@iastate.edu}

\author{J. Pablo Mu{\~{n}}oz}
\affiliation{%
  \institution{Intel Corporation}
  \city{Santa Clara, CA}
  \country{USA}}
\email{pablo.munoz@intel.com}

\author{Ali Jannesari}
\affiliation{%
  \institution{Iowa State University}
  \city{Ames, IA}
  \country{USA}}
\email{jannesari@iastate.edu}


\begin{abstract}

Concerned with user data privacy, this paper presents a new federated learning (FL) method that trains machine learning models on edge devices without accessing sensitive data. Traditional FL methods, although privacy-protective, fail to manage model heterogeneity and incur high communication costs due to their reliance on aggregation methods. To address this limitation, we propose a resource-aware FL method that aggregates local knowledge from edge models and distills it into robust global knowledge through knowledge distillation. This method allows efficient multi-model knowledge fusion and the deployment of resource-aware models while preserving model heterogeneity. Our method improves communication cost and performance in heterogeneous data and models compared to existing FL algorithms. Notably, it reduces the communication cost of ResNet-32 by up to 50\% and VGG-11 by up to 10$\times$ while delivering superior performance.

\end{abstract}


\maketitle

\section{Introduction}

Federated learning (FL) has emerged as a novel machine learning paradigm for distributed clients to participate in the collaborative training of a centralized model. FL brings model synchronous training at the network edge, where devices (e.g., mobile phones and IoT devices) extract the knowledge on the private-sensitive training data and then upload the learned models to the cloud for aggregation. FL stores user data locally and restricts direct access to it from cloud servers; thereby, this paradigm not only enhances privacy-preserving but also introduces several inherent advantages, including model accuracy, cost efficiency, and diversity. With the massive demand for data in today's machine learning models and the social considerations of artificial intelligence (e.g., privacy and security \cite{yang_federated_2019,curzon21privacy}), federated learning has great potential and a role in counterpoising this trade-off.

Federated learning has already shown its potential in practical applications, including health care \cite{healthcare}, environment protection \cite{environment}, and electrical vehicles \cite{electricvehicle} to name a few applications. 
Intel, Google, Apple, and NVIDIA are using FL for their applications nowadays (e.g., Intel's OpenFL \cite{openfl_citation}, Google's Keyboard~\cite{yang2018keyboard,chen2019keyboard}, Apple's Siri~\cite{apple2019siri,paulik2021siri}, NVIDIA's medical imaging~\cite{li2019nvidia}). In consequence, designing efficient FL models and deploying them effectively and fairly on edge devices is crucial for improving the performance of edge computing in the future.

\begin{figure*}[t]
\centering

\centerline{\includegraphics[width=0.8\textwidth]{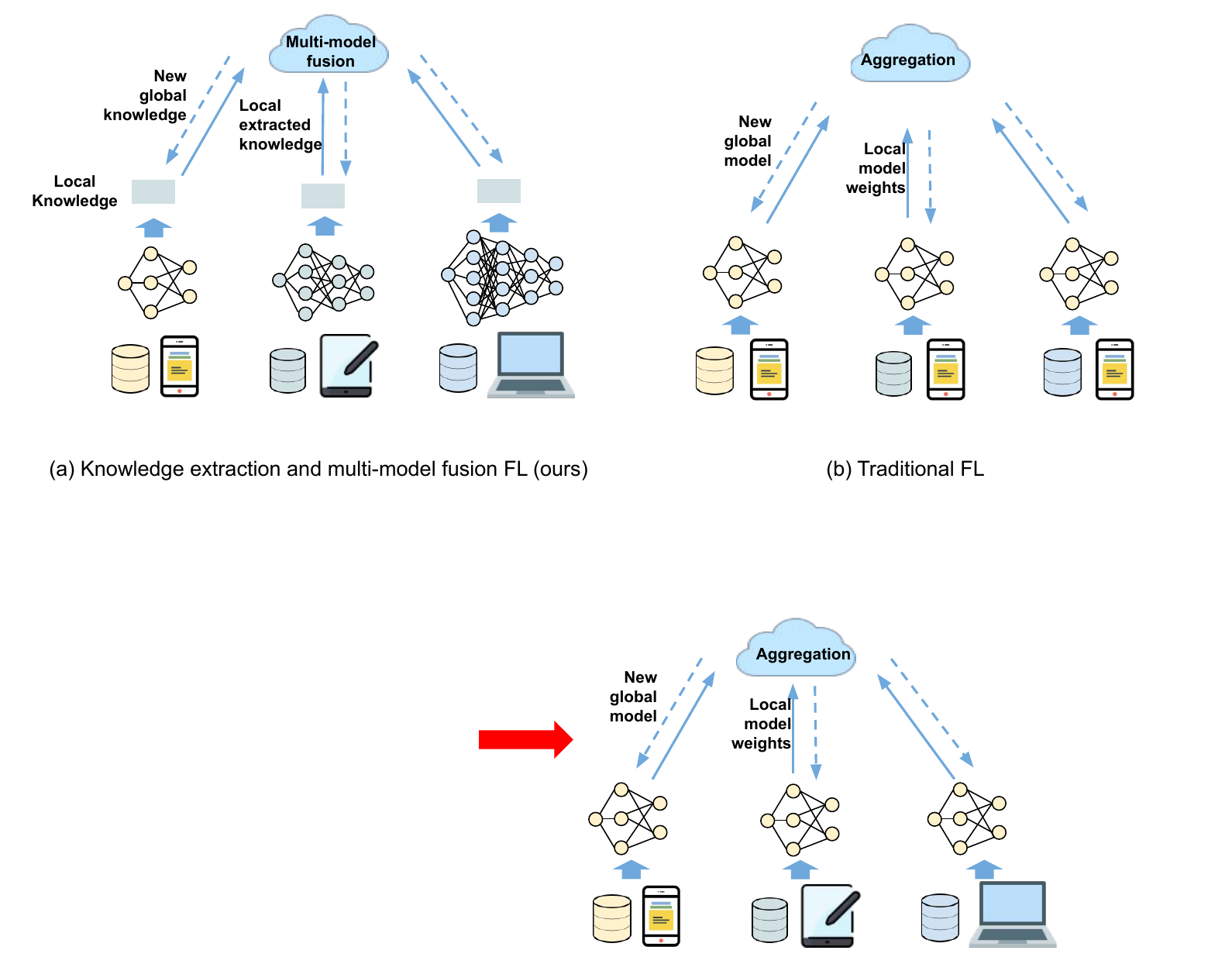}}

\caption{(a) The proposed knowledge extraction and multi-model fusion FL method builds global knowledge by extracting local knowledge from different models using the corresponding computing power device and fuses it in, then transfers the local knowledge to the edge. It can thus be aware of resource constraints and serve as a robust general-purpose FL solution for practical applications. (b) In contrast, traditional FL produces aggregation of local model weights and distributes the global model to edges. It treats edge devices with the same model and computing power and has complex communication costs.}

\label{fig:FL_compare}
\end{figure*}

Traditional FL, like FedAvg~\cite{mcmahan2017fedavg} in Figure \ref{fig:FL_compare} (b), broadcasts global model parameters to selected edge devices, then averages trained local model parameters based on local data to update the global model. This iterative process, while effective, faces limitations~\cite{Michieli_2021_CVPR}. Firstly, model weight sharing between network edges and servers introduces significant communication overhead. Secondly, increasing computational and memory demands of AI models, combined with the heterogenous computing power of edge devices, complicate model deployment on resource-constrained devices. Additionally, real-world local data often exhibits imbalance or non-independence (Non-IID), which can result in training failures in decentralized situations. Lastly, over-parameterized deep learning models can cause overfitting when aggregating heterogeneous local models~\cite{jiang_improving_2019,nishio_client_2019}, leading to high learning and prediction variance. Hence, existing FL methods may result in an unfair and ineffective global model.

We propose a resource-aware Federated Learning (FL) approach, \textbf{Fed}erated learning using \textbf{K}nowledge \textbf{E}xtraction and \textbf{M}ulti-model fusion~(\proj) as illustrated in Figure \ref{fig:FL_compare} (a), to overcome previous FL limitations. In \proj, each local client is trained with a server-side knowledge network. After local training, each local model's knowledge is distilled into a tiny-size neural network via deep mutual learning, allowing robust global knowledge acquisition from the server. We then ensemble the tiny neural networks, distill this to global knowledge and transfer it to the client for further learning. This reduces communication costs as only the tiny network is exchanged between edge and cloud, during training and inference.

\proj offers several advantages: It prevents large edge models' over-parameterization by distilling client knowledge before server aggregation. It cuts communication costs significantly by exchanging distilled tiny-size networks instead of the original large models. By ensembling knowledge from edges, the global model is strengthened, reducing overfitting and variance risks, and improving FL generalization. Additionally, it considers resource limitations, using multi-model fusion to deploy models fairly on edge devices, making FL more realistic.

Our experiments on non-IID data settings and heterogeneous client models show FedKEM significantly reduces communication cost and achieves better performance using fewer communication rounds.

\section{Related work}
\label{sec:relat}
\subsection{Federated Learning}

\begin{figure}[t]
\centering

\includegraphics[width=1.0\columnwidth]{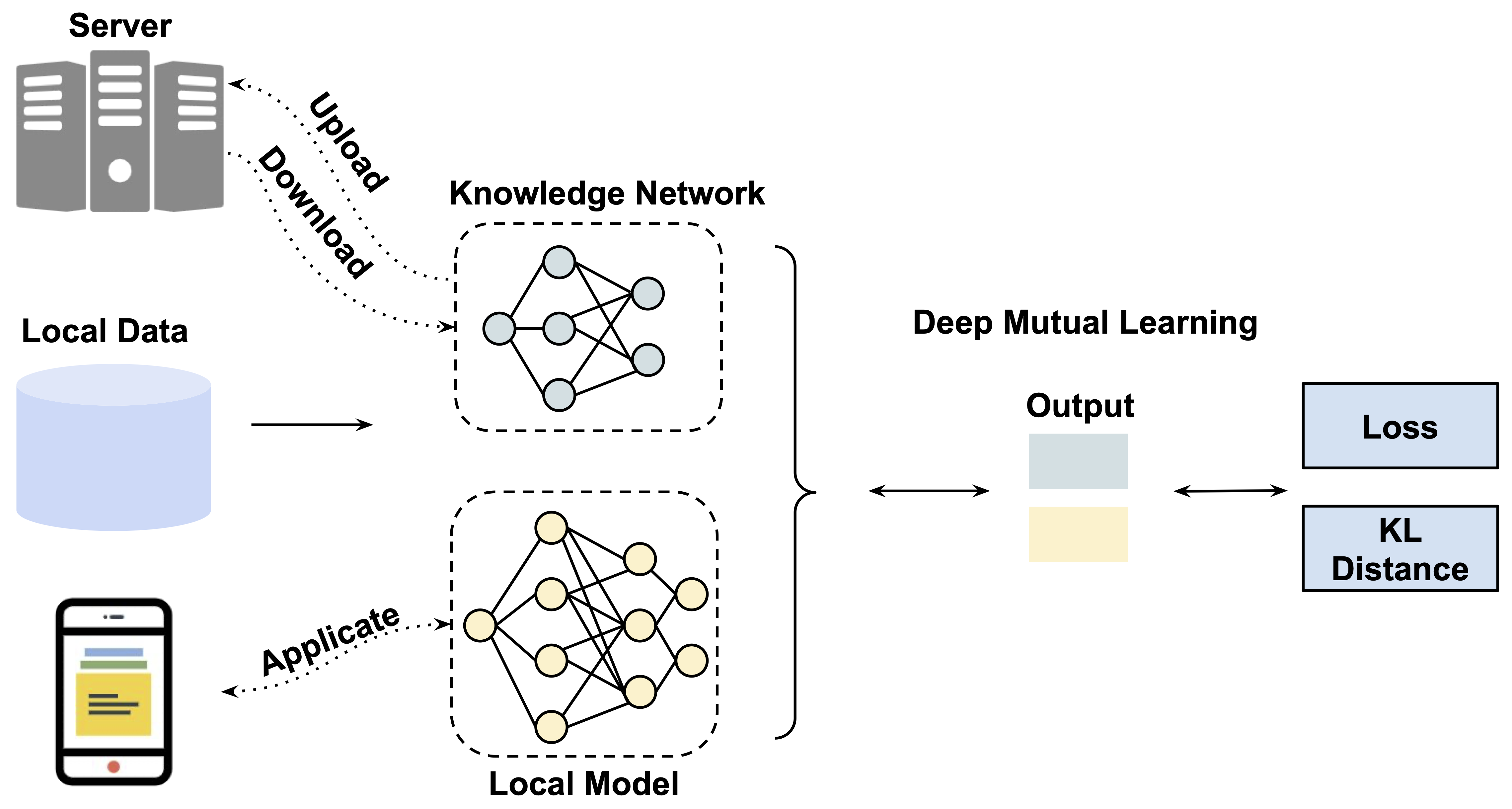}
\caption{Local updates. The client downloads the knowledge network from the server, trains mutually with the local model, extracts the updated knowledge network, and uploads it to the server.}

\label{fig:local_update}
\end{figure}

FedAvg \cite{mcmahan2017fedavg} is the original implementation for training decentralized data and preserving privacy in FL. Based on FedAvg, numerous variants have been proposed to optimize FL \cite{li2020fedprox,wang2020fednova,karimireddy2020scaffold}, especially to track the heterogeneity issue in FL \cite{karimireddy2020scaffold,li2020convergence_noniid,yu2022rafl}. For example, FedProx \cite{li2020fedprox} is proposed to improve local client training by adding a proximal term to the local loss, FedNova \cite{wang2020fednova} introduces weight modification to avoid gradient bias by normalizing and scaling local updates, SCAFFOLD \cite{karimireddy2020scaffold} corrects the update direction to prevent client drift problem by maintaining drift variates. It is worth mentioning that these FL methods aggregate single-model weights for server from edges and incurs extra communication overhead. Other works \cite{basak_2020,elkordy_2020,yuan_2020,yu23dynamicpruning} optimize the communication cost but have no consideration of the computation power heterogeneity of edge devices. Unlike prior works, we aggregate knowledge from the edge and only communicate these tiny knowledge networks between the edge and the server.

Another line in FL is personalized FL, which focuses on the problem of statistical heterogeneity. Personalized FL aims to personalize the global model for each client in FL and find how to develop improved personalized models that can benefit a large majority of clients \cite{kulkarni_survey_2020}.  SPATL~\cite{yu20spatl} introduces a knowledge transfer local predictor
that transfers the shared encoder to each client. It further leverages network pruning~\cite{yu2022gnnrl,yu2021agmc} to select salient parameters in communication, only communicating selected parameters between the server and clients. Although we have the same consideration of device heterogeneity (memory storage and computation power), data heterogeneity (non-IID data), and model heterogeneity (model structure and size), we focus on how to extract knowledge from different types of models and their corresponding training devices to build robust global knowledge.

\subsection{Ensemble Learning and Knowledge Distillation in Federated Learning}

Knowledge distillation is first introduced as a model compression technique for neural networks to transfer knowledge from a large teacher model to a small student model \cite{modelcompression,hinton_distilling_2015}. Ensemble learning is a promising technique to combine several individual models for better generalization performance \cite{ganaie_ensemble_2022}. In light of these two ideas, such methods \cite{you_learning_2017,fukuda_efficient_2017,tung_similarity-preserving_2019,park_feature-level_2020,asif_ensemble_2020} purpose efficient knowledge distillation with an ensemble of teachers to further improve the student performance.

Recently, ensemble learning and knowledge distillation have emerged as an effective approach to address heterogeneity issues, resource-constrained edge devices, and communication efficiency in FL~\cite{li_fedmd_2019,he_group_2020,seo_federated_2020,sui-etal-2020-feded,lin_ensemble_2020}. For example, Fed-ensemble~\cite{sui-etal-2020-feded} ensembles the prediction output of all client models; FedKD~\cite{wu_fedkd_2021} proposes an adaptive mutual distillation to learn a student and a teacher model simultaneously on the client side; FedDF~\cite{lin_ensemble_2020} distills the ensemble of client teacher models to a server student model. In contrast, our approach utilizes knowledge distillation to encode the ensemble knowledge from clients into global knowledge. The novelty of our approach is that \proj can extract the local model knowledge that is being learned from the global knowledge into a tiny-size neural network by deep mutual learning~\cite{zhang_deep_2018}, and then ensembles the tiny-size neural networks for multi-model fusion.


\section{Methodology}
\label{sec:method}

\begin{figure}[t]
\centering

\includegraphics[width=1.0\columnwidth]{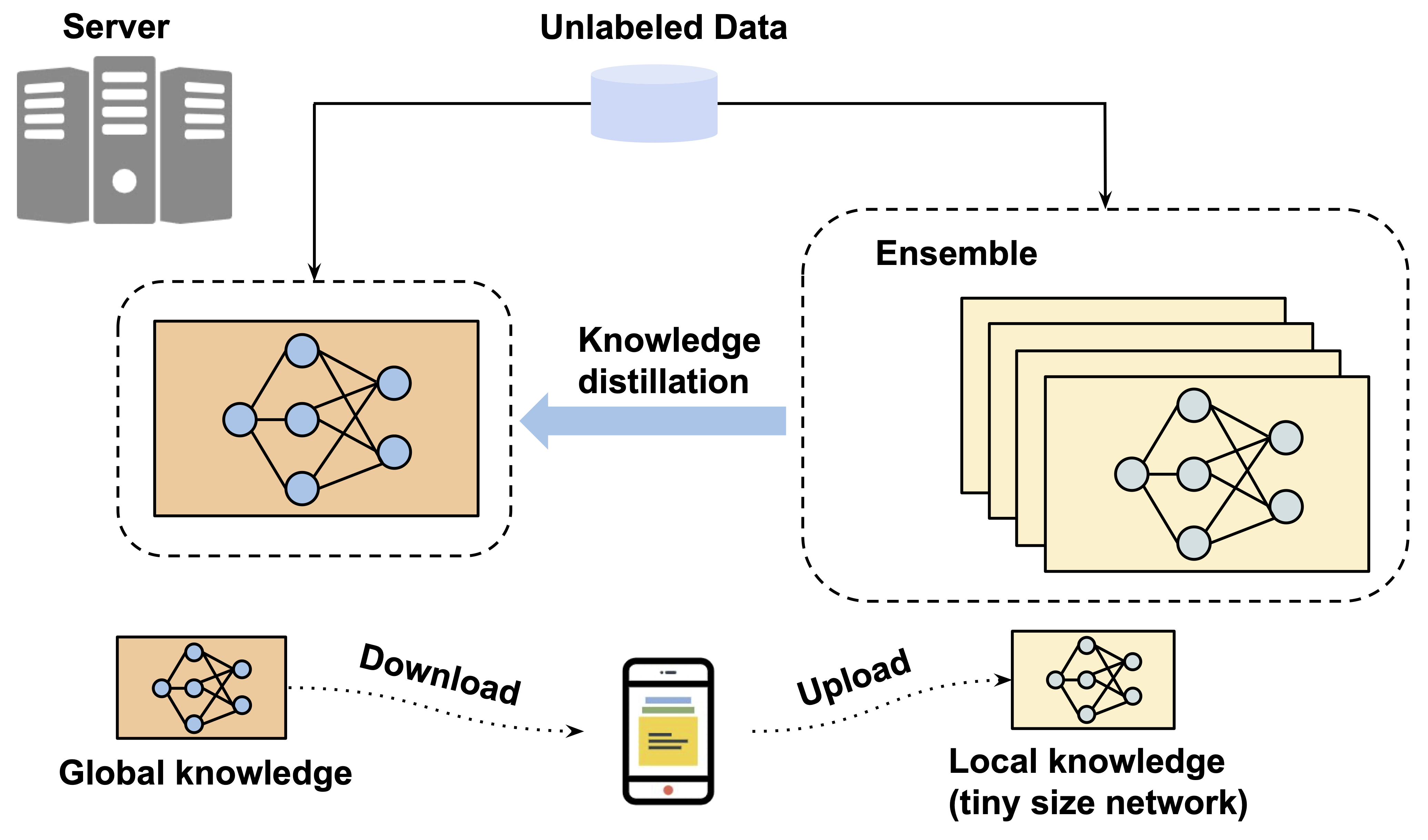}
\caption{Cloud updates. The cloud collects local knowledge (tiny-size network) from clients, ensembles all tiny-size networks, distills into global knowledge and transfers it to clients.} 

\label{fig:cloud_update}
\end{figure}

Figure \ref{fig:local_update} shows the local updates of \proj and Figure \ref{fig:cloud_update} shows the cloud updates of \proj. In local updates, the client first downloads the knowledge network from the server, then mutually trains the knowledge network with the local model to extract an updated knowledge network, and finally uploads it back to the server. In cloud updates, the server first collects local knowledge (tiny-size network) uploaded from clients, then ensembles all the tiny-size networks, distills them into global knowledge, and finally transfers it to clients. In this section, we will explain our approach in detail.

\subsection{Knowledge Extraction using Deep Mutual Learning}
Traditional FL and its variants~\cite{li2020fedprox,wang2020fednova,karimireddy2020scaffold} keep the model up-to-date by sharing model weights/gradients between server and edge clients.
Simply aggregating weights might raise unexpected training failures.  
We aim to fuse the local model’s knowledge to keep the model updated. Hence, we use deep mutual learning~\cite{zhang_deep_2018}, to extract the knowledge.

The key idea of deep mutual learning is to train multiple neural networks~(NNs) synchronously while minimizing the Kullback Leibler~(KL) divergence among the output of the networks.  In other words, the KL divergence evaluates the similarity of two distributions. By minimizing the KL divergence among the NNs, NNs can learn knowledge from each other. 
Therefore, to extract knowledge from the local model, we introduce a knowledge network~(a tiny size network compared to the local model) locally and optimize the knowledge network and local model simultaneously using deep mutual learning.

To intuitively explain the knowledge extraction process, we explain it in an image classification task. Formally, in an edge client, we have a local model $\theta$ and a knowledge network~(tiny size network) $\theta_g$. 
First, to update the local model $\theta$. For any input batch of data $x$, we calculate the cross-entropy loss of predictions and ground truth as equation~\ref{eq:ce}.
\begin{equation}
\label{eq:ce}
    L_c = -\sum_{i=1}^M y^T log(\sigma(\theta(x_i))),
\end{equation}
where $M$ is the mini-batch size, $y$ is the ground truth label, and $\sigma$ is the softmax function.

Then, we compute the KL divergence from $\theta(x)$ to $\theta_g(x)$ as equation~\ref{eq:kl1}.
\begin{equation}
    \label{eq:kl1}
    D_{KL}(\theta_g||\theta) = \sum_{i=1}^M \sigma(\theta_g(x)^T) log(\frac{\sigma(\theta_g(x))}{\sigma(\theta(x))}),
\end{equation}
where we do element-wise division on $\frac{\sigma(\theta_g(x))}{\sigma(\theta(x))}$.

We update $\theta$ by using the total loss as equation~\ref{eq:loss1}.
\begin{equation}
    \label{eq:loss1}
    L_{\theta} = L_c + D_{KL}(\theta_g||\theta)
\end{equation}
Similar steps are followed to update $\theta_g$.

\subsection{Local Updates Through Deep Mutual Learning}
As depicted in Figure~\ref{fig:local_update}, we mutually train the local model and knowledge network, transmitting the knowledge network back and forcing it to update the global knowledge.
Knowledge extraction and communication allow edge clients to deploy resource-aware models for the application while keeping communication efficiency and model heterogeneity.
Algorithm~\ref{alg:local} shows the local update process.

\begin{algorithm}[tb]
\caption{\proj Edge Client Update}
\label{alg:local}

\SetKwInOut{KwIn}{Input}
\SetKwInOut{KwOut}{Output}

\KwIn{Number of local epochs $\mathcal{E}$, learning rate $\eta$}

\KwOut{Deploy local model $\theta$ and return knowledge model $\theta_g$}


$\theta \leftarrow$ local deployed model

$\mathcal{B} \leftarrow$ split local dataset into batches 

\For{$e=1,2,\ldots, \mathcal{E}$} {
\For{each batch $b\in \mathcal{B}$}{

\tcc{Perform Deep Mutual Learning}

$\theta \leftarrow \theta - \eta \triangledown(\mathcal{L}(\theta;b) + D_{KL}(\theta_g||\theta)) $ 

$\theta_g \leftarrow \theta_g - \eta \triangledown(\mathcal{L}(\theta_g;b) + D_{KL}(\theta||\theta_g)) $ 

}
}

Deploy $\theta$ on local application.

\KwRet{$\theta_g$}

\end{algorithm}

\subsection{Multi-model Knowledge Fusion}
In \proj, we provide two model fusion methods for server fusion of the knowledge from the edge. The first one is similar to the traditional FL in that we aggregate the weight. Second, inspired by FedDF, we ensemble all received client models and distill the ensemble knowledge into a global knowledge network. In this section and the experiments, we mainly focus on ensemble the client's knowledge. However, \proj can also use traditional fusion methods, such as FedAvg and SCAFFOLD to aggregate the model.

We define the ensemble model as $\Theta = \{\theta_g^k\}_{k\in S}$, where the $\theta_g^k$ is the $k^{th}$ client's knowledge network and the $S$ is the set of clients that communicate with the server in current communication round. Then we distillate the knowledge of ensemble $\Theta$ to a global knowledge network $\theta_g$ by using unlabeled data, generative data, or public data in the server. The distillation loss for $\theta_g$ is defined in equation~\ref{eq:dist}.

\begin{equation}
\label{eq:dist}
    L_d = D_{KL}(\Theta, \theta_g) 
\end{equation}

The server update process is shown in the Algorithm~\ref{alg:server}.

\begin{algorithm}[tb]
\caption{\proj Server Update}
\label{alg:server}

\SetKwInOut{KwIn}{Input}
\SetKwInOut{KwOut}{Output}

\KwIn{Number of communication rounds $T$, total number of clients $N$, client sample ratio $f$, learning rate $\eta$}

\KwOut{Knowledge model $\theta_g$}


Initialize $\theta_g$.

\For{$t = 1, 2,\ldots, T$} {

$m \leftarrow$ max$(\lfloor Nf \rfloor, 1)$

$S \leftarrow$ random set of $m$ clients

\For{each client $k \in S$ in parallel}{

\textbf{Communicate with client $k^{th}$}

$\theta_g^k \leftarrow$ \textbf{ClientUpdate}$(\theta_g)$
}
}

$\Theta = Ensemble(\{\theta_g^k\}_{k\in S})$

$\theta_g \leftarrow \theta_g - \eta \triangledown(\mathcal{L}_d(\Theta,\theta_g)) $ 

\KwRet {$\theta_g$}
\end{algorithm}

\subsection{Ensemble Knowledge}
In \proj, we investigate three ensemble strategies, i.e., max logits, average logits, and majority vote. We adopt the max logits as the ensemble strategy since the max logits get the best results in practice.
For a given input instance $x$, the ensemble model is obtained by the following equation:
\begin{equation}
    \Theta(x) = Ensemble_{Max}( \{\theta_g^k(x)\}_{k\in S} ),
\end{equation}
where the $Ensemble_{Max}$ compares all output vectors and returns a new vector containing the element-wise maxima.

\section{Experiments}
\label{sec:exp}

\begin{figure*}
 \begin{center}

\centerline{\includegraphics[width=0.9\linewidth]{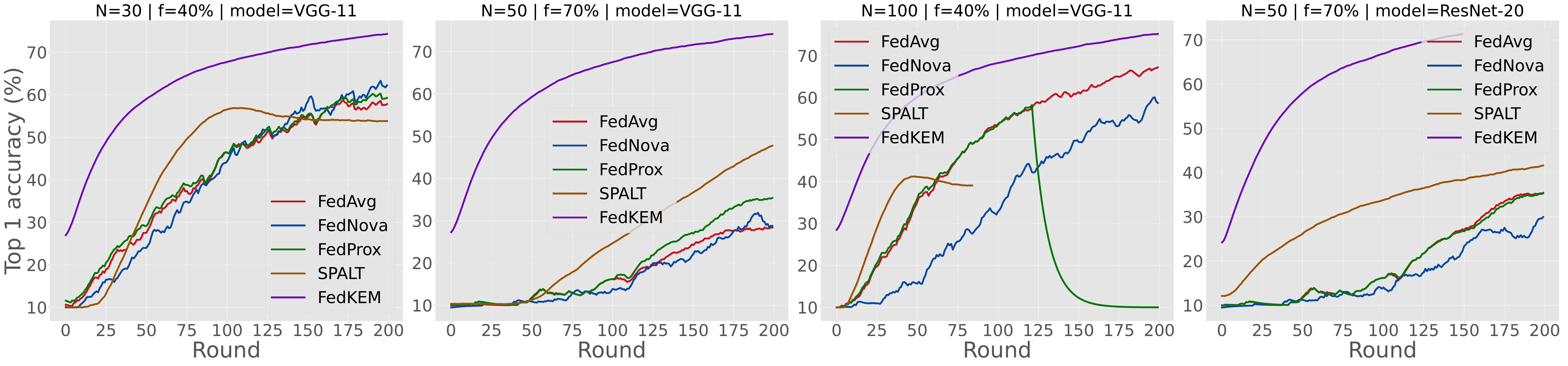}}
\centerline{\includegraphics[width=0.9\linewidth]{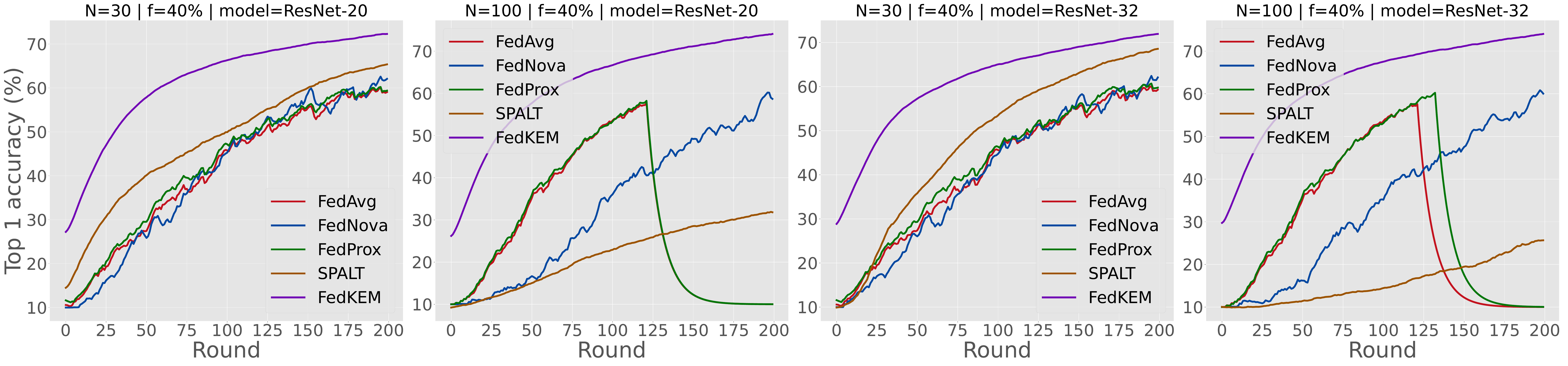}}
\caption{Comparison of \proj with FedAvg~\cite{mcmahan2017fedavg}, FedNova~\cite{wang2020fednova}, FedProx~\cite{li2020fedprox}, SPATL~\cite{yu20spatl}: the top-1 test accuracy vs. communication rounds. $N$ is the total number of clients, $f$ is the client sample ratio.  The model type is the network at local devices and also the network used for communication in the case of the baseline methods). The knowledge network is ResNet-20 in all cases for \proj.}
\vspace{-1em}
\label{fig:learn_effi}
\end{center}
 \end{figure*}

We conduct comprehensive experiments to evaluate the performance of \proj.
We separated our experiments into three sections: learning efficiency, communication cost, and multi-model federated learning. In addition, we performed an ablation study with different ensemble methods for \proj.

\subsection{Experimental Setup}

\textbf{Datasets and models.} We run experiments on CIFAR-10~\cite{krizhevsky2009cifar}. Various computer visions are used such as VGG-11~\cite{simonyan2015vgg}, EfficienNet~\cite{tan2019efficientnet},  MobileNet~\cite{howard2017mobilenet}, ResNet~\cite{he2016resnet} family with three variants: ResNet-20, ResNet-32 and ResNet-44. \\
\textbf{Clients and sample rate.} We use three different settings: 30 clients with 40\% sample rate, 50 clients with 70\% sample rate, and 100 clients with 40\% sample rate. In each case respectively, 12, 25, and 40 clients are participating in each communication round .\\
\textbf{Data partition.} We implement federated learning following the non-IID benchmark setting \cite{li2022NonIIDBench}. The partition scheme follows a Dirichlet distribution with each client being assigned a sample proportion of each label based on $\beta$.
The lower $\beta$ corresponds to higher heterogeneity. We chose $\beta = 0.1$ in this experiment. 
\textbf{Training procedure.} We train for 200 communication rounds with 10 local epochs. Therefore, the total number of local updates is 2000. The batch size is 64. Stochastic gradient descent is utilized for optimization. The total sample size of testing data is 10000. \\
\textbf{Evaluation.} We train the local models on their respective training data for each local update. The global model is then evaluated on the whole testing data set of CIFAR-10 at the end of each communication round.\\
\textbf{Baselines.} We compare \proj with state-of-the-art (SoTA) FL algorithms, including FedAvg~\cite{mcmahan2017fedavg}, FedProx~\cite{li2020fedprox}, FedNova~\cite{wang2020fednova} and SPATL~ \cite{yu20spatl}.

\subsection{Learning Efficiency}

In our evaluation, we assess \proj's learning efficiency and optimization, particularly analyzing the correlation between communication rounds and the target model's accuracy. The universal models trained for the baselines include VGG-11, ResNet-20, and ResNet-32, while ResNet-18 is used as the knowledge network for \proj, due to its status as a commonly utilized, minimal model in the ResNet family. 

Figure~\ref{fig:learn_effi} reveals that \proj delivers superior results in most benchmark settings compared to robust FL baselines, exhibiting a stable training process. It significantly surpasses baselines in handling over-parameterized networks like VGG-11 and is particularly adept at managing heterogeneous settings. For example, with 30 clients, \proj achieves 70\% accuracy after 110 rounds, while all baselines fail to reach this accuracy even after 200 rounds. This gap widens with 50 clients, where baselines fail to attain 50\% accuracy after 20000 total local updates.

These results demonstrate \proj's advantages in stability and consistency, particularly in high heterogeneity FL environments. However, SPATL, although performing well with 30 and 50 clients, does not compete effectively with other baselines when the number of clients increases to 100.

\begin{figure}
 \begin{center}

\centerline{\includegraphics[width=1.0\linewidth]{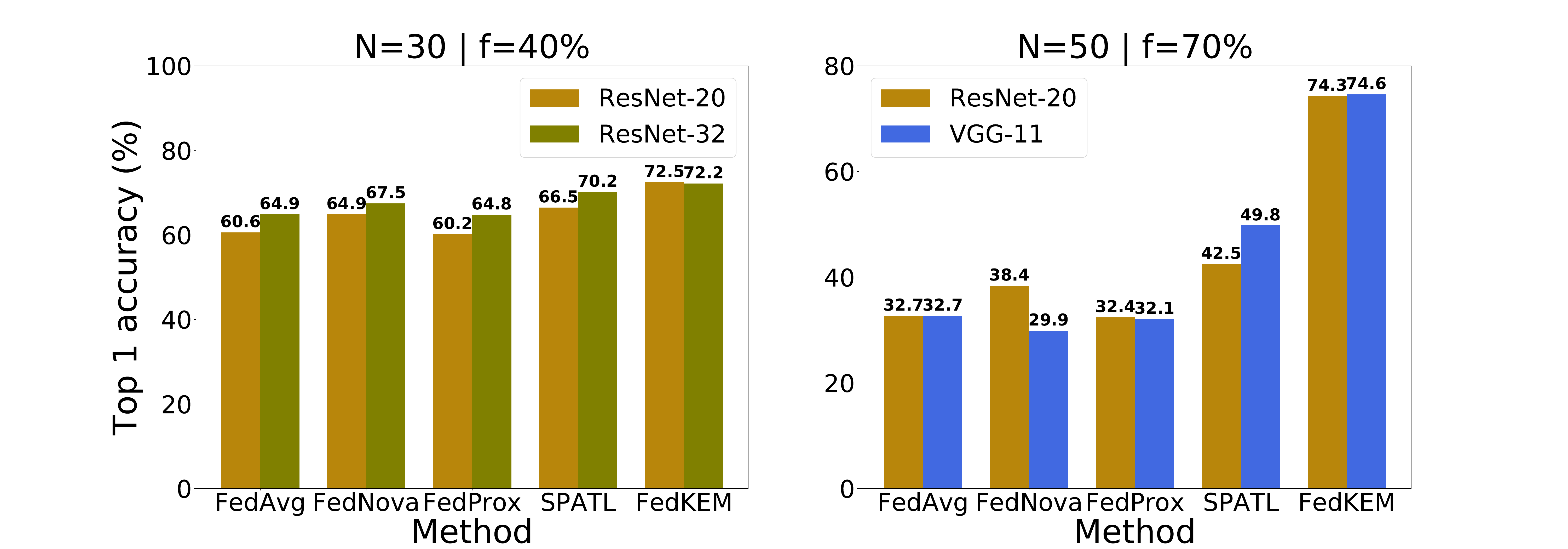}}
\vspace{-1em}
\caption{Comparison of \proj with FedAvg~\cite{mcmahan2017fedavg}, FedNova~\cite{wang2020fednova}, FedProx~\cite{li2020fedprox}, SPATL~\cite{yu20spatl}: the convergence accuracy. $N$ is the total number of clients, $f$ is the client sample ratio.  The model type is the network at local devices and also the network used for communication in the case of the baseline methods). The knowledge network is ResNet-20 in all cases for \proj. The higher the better.}
\vspace{-1.5em}
\label{fig:converged_acc}
\end{center}
 \end{figure}

We observe that the knowledge model used in \proj doesn't significantly impact its performance. It consistently maintains stable training processes and high final convergence (over 70\%) across all settings. The final converged accuracies for 30 and 50 clients are documented in Figure~\ref{fig:converged_acc}. Notably, \proj avoids gradient explosions during training, an issue observed in other methods~\cite{yu20spatl}.

Baseline methods primarily concentrate on parameter and gradient aggregation for model fusion in the cloud. However, their aggregation approach, such as FedAvg's weighted averaging, can introduce biases. This is primarily due to the contribution of individual edge models to the FL system being a black box.

On the contrary, \proj uses ensemble distillation for model fusion. This approach generalizes heterogeneous edge models effectively, guiding the model towards an optimal direction, resulting in a much more stable optimization process across various non-IID FL settings.

\begin{figure}
 \begin{center}


\centerline{\includegraphics[width=1.0\linewidth]{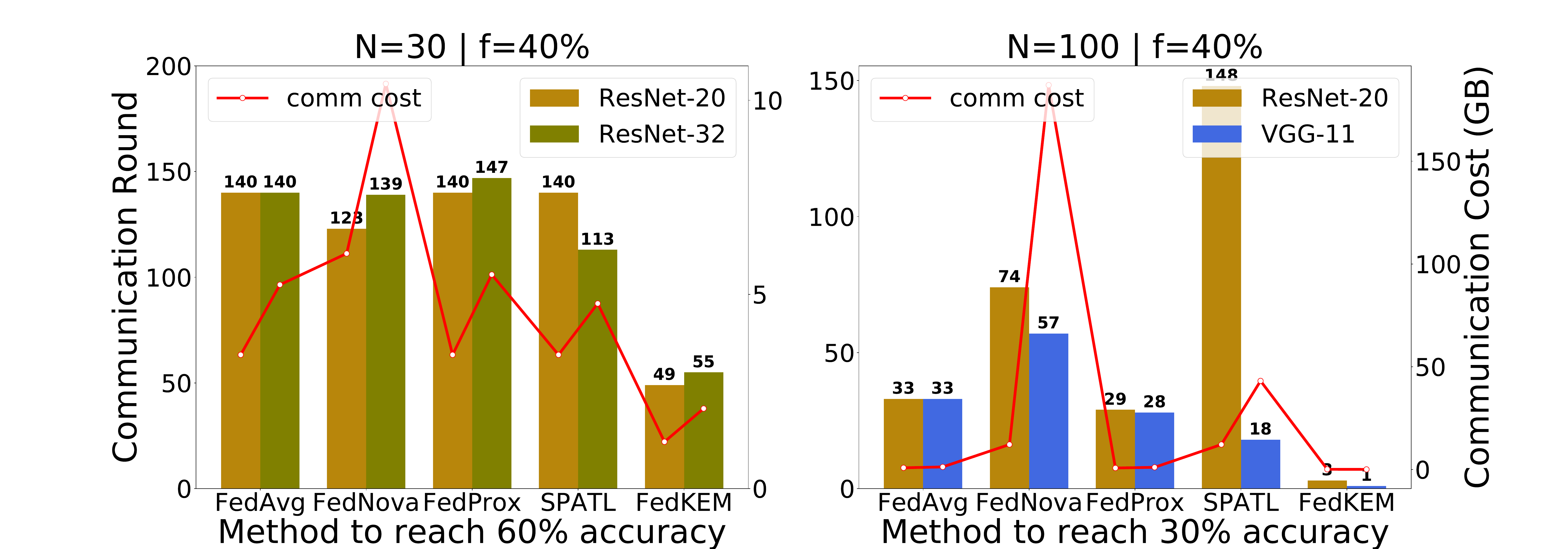}}
\vspace{-1em}
\caption{Comparison of \proj with FedAvg~\cite{mcmahan2017fedavg}, FedNova~\cite{wang2020fednova}, FedProx~\cite{li2020fedprox}, SPATL~\cite{yu20spatl}: the communication rounds and overhead needed to read 60\% and 30\% top-1 accuracy. $N$ is the total number of clients, $f$ is the client sample ratio.  The model type is the network at local devices and also the network used for communication in the case of the baseline methods). The knowledge network is ResNet-20 in all cases for \proj.}
\vspace{-1.5em}
\label{fig:comm_cost}
\end{center}
 \end{figure}

\begin{figure*}
 \begin{center}

\centerline{\includegraphics[width=0.8\linewidth]{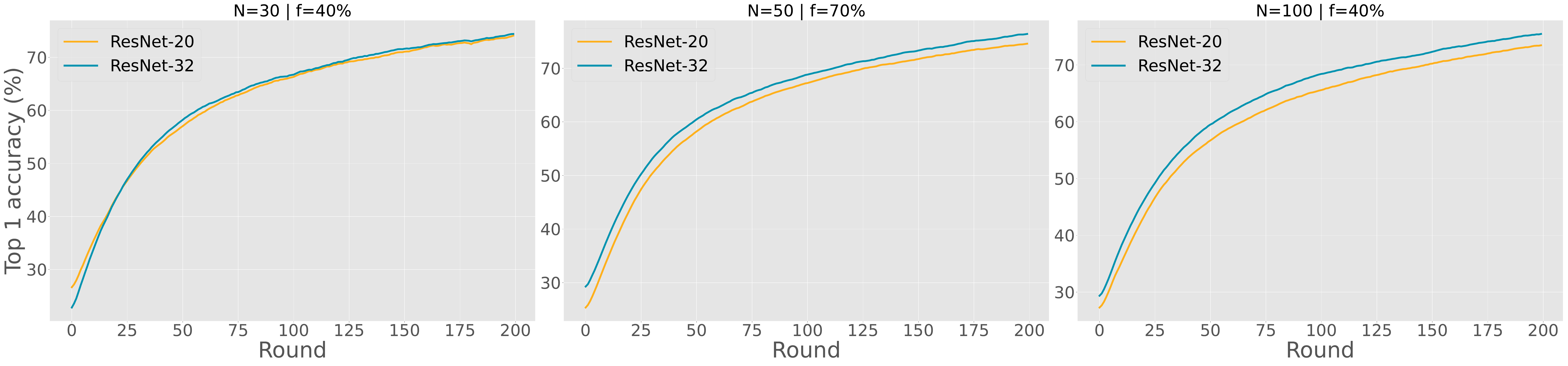}}
\vspace{-1.em}
\caption{Comparison of \proj with multi-model: the top-1 test accuracy vs. communication rounds. $N$ is the total number of clients, $f$ is the client sample ratio, and the model type is the network used for communication in the baselines. ResNet-20 and ResNet-32 are used as the knowledge network.}
\vspace{-1em}
\label{fig:multi_model}
\end{center}
 \end{figure*}

\begin{table*}
\centering
\caption{Models distribution at local clients in the multi-model setting.}
\label{tab:config_mm}
\begin{tabular}{ccc}
\hline
\textbf{} & \textbf{knowledge network} & \textbf{\# of \{ResNet-20, ResNet-32, ResNet-44, VGG-11\}} \\ \hline
\multirow{2}{*}{30 clients}  & ResNet-20 & \{6, 5, 10, 9\}    \\
                             & ResNet-32 & \{0, 8, 13, 9\}    \\ \hline
\multirow{2}{*}{50 clients}  & ResNet-20 & \{13, 10, 14, 13\} \\
                             & ResNet-32 & \{0, 13, 20, 17\}  \\ \hline
\multirow{2}{*}{100 clients} & ResNet-20 & \{26, 24, 28, 22\} \\
                             & ResNet-32 & \{0, 31, 34, 35\}  \\ \hline
\end{tabular}%
\end{table*}

\subsection{Communication Efficiency}
In \proj, a key attribute is that we introduce a knowledge network independent of the edge models. Hence, it only communicates the tiny size knowledge network through training and inference, resulting in lower communication costs than SoTA FL algorithms that use models for aggregation and fusion. 
For example, FedNova can achieve stable training but costs double the average communication cost compared to FedAvg as a result of sharing the extra gradient information. We evaluate the communication cost in two ways. First, we train all the models to a target accuracy and calculate the communication cost. Second, we train all the models to converge and then compare the communication cost for each FL algorithm. The communication cost is represented by:
\begin{equation}
\label{eq:cost}
    N \times f \times T \times
    \text{round cost}, 
\end{equation}
where the round cost is $2 \times$ size of the exchanged model (for downloading and uploading the global model in case of baselines or the knowledge model in case of \proj).

Figure~\ref{fig:comm_cost} shows the results of communication rounds needed to reach target accuracy. \proj requires much fewer rounds to reach 60\% accuracy compared to other baselines. For example, with $N=30$ and ResNet-20 as the knowledge network, \proj needs 49 rounds while the next best, FedNova requires 123 rounds.

When factoring in the communication cost, the difference is even more renounced. 
With the model size of VGG-11, ResNet-20, ResNet-32 and ResNet-44 given as 21, 1.05, 1.6, and 2.7 Mb respectively, we can calculate the transmission overhead of each method using equation ~\ref{eq:cost}. For instance, with ResNet-32 as the knowledge network, the communication burden of SPATL is more than double that of \proj to reach the same 60\% accuracy (4.76 GB compared to 2.06 GB).

The reduction in communication cost is even more notable when an over-parameter network like VGG-11 is concerned with higher heterogeneity (bigger $N$). With $N=100$ and ResNet-20 as the knowledge network, \proj needs 0.074 GB for transferring while FedProx needs 0.71 GB to get the same result, a one-order of-magnitude cutback.



\subsection{Multi-Model Federated Learning}

Federated learning systems often face challenges in data and resource heterogeneity. While existing FL works primarily focus on addressing data heterogeneity and reducing training performance overhead, resource heterogeneity remains a significant challenge. A uniform model deployed across all resource-heterogeneous edge clients may limit the system's computational overhead due to resource-poor clients.

\proj tackles these challenges through knowledge extraction, enabling multi-model deployment on heterogeneous edge devices. This allows models to be deployed more effectively to edge clients based on their computational resources, rather than sharing an identical model after optimization as in traditional FL methods. Depending on clients' memory and Multiply-Add accumulation (MACs), we can allocate suitable models for better utilization.

We evaluated \proj's performance on multi-model deployment using ResNet-20/32/44 in the same FL system, updating the multi-model edge clients with ResNet-20 as the knowledge model. This approach is beneficial when clients have different computational capabilities, as it enables local network customization. \proj showed stable training and quickly achieved high accuracy despite different client models.

In Figure~\ref{fig:multi_model}, we compare \proj's multi-model performance with ResNet-20 and ResNet-32 as the knowledge network, choosing the smallest to minimize communication overhead. The results showed comparable accuracy (over 70\%) to training a single universal model. The performance difference between ResNet-20 and ResNet-32 as knowledge networks became more pronounced as the heterogeneity (N) increased.

\section{Conclusion}
\label{sec:con}


This paper presents \proj, a novel federated learning paradigm that addresses the challenges of limited resources and computing power heterogeneity in edge devices. \proj uses a compact network architecture that is trained locally to extract knowledge from the local model and integrate global knowledge. This "tiny-size" network is then sent to a central service for multi-model fusion and global knowledge distillation. Experiment results reveal \proj's superiority over other federated learning algorithms in accuracy, efficiency, and stability, underlining its potential as a scalable solution for federated learning issues. Future research will aim to enhance multi-model fusion efficiency and explore the approach's applicability in other machine learning areas. Ultimately, we strive to position \proj as a leading solution for federated learning problems.

\printbibliography

\end{document}